\documentclass[prd,amsmath]{revtex4}
\usepackage{graphicx}
\usepackage{hyperref}
\usepackage{bm}
\usepackage{natbib}

\begin{document}
\title{Magnetic Helicity in Sphaleron Debris}

\author{Yi-Zen Chu, James B. Dent and Tanmay Vachaspati}
\affiliation{Physics Department, Arizona State University, Tempe, AZ 85287}

\begin{abstract}
We develop an analytical technique to evaluate the magnetic helicity in
the debris from sphaleron decay. We show that baryon number production
leads to left-handed magnetic fields, and that the magnetic helicity
is conserved at late times. Our analysis explicitly demonstrates the
connection between sphaleron-mediated cosmic baryogenesis and cosmic
magnetogenesis.
\end{abstract}

\maketitle

\section{Introduction}
\label{sec:intro}

A cosmic phase transition in which magnetic monopoles are produced
necessarily produces cosmic magnetic fields, as these are sourced by
the monopoles. The electroweak model also contains magnetic monopoles
\cite{Nambu:1977ag} that can source magnetic fields
\cite{Vachaspati:1991nm,Vachaspati:1994xc}. However, a quantitative
estimate of the magnetic field is difficult to obtain because
electroweak monopoles are confined to electroweak antimonopoles by
Z-strings \cite{Vachaspati:1992fi}, and estimates of the
monopole density based on topological considerations cannot be applied.
Fortunately there is a way to sidestep this difficulty because
magnetic monopole-antimonopole pairs, in the guise of ``sphalerons''
\cite{Manton:1983nd}, play a crucial role in the violation of baryon
number in the electroweak model \cite{Vachaspati:1994ng}. Hence
monopole-antimonopole pairs are a common factor in baryogenesis
and magnetogenesis. This fact allows one to estimate the helicity of
the magnetic field by relating it to the baryonic density
\cite{Cornwall:1997ms,Vachaspati:2001nb}. The connection between
baryons and magnetic fields may provide an opportunity for tests of
fundamental particle physics and cosmology at the electroweak epoch
via observations of the topological and spectral properties
of a primordial magnetic field.

To test the production of magnetic fields during baryon number
violating processes, sphaleron decay was studied in \cite{Copi:2008he}
(also see \cite{GarciaBellido:2003wd})
by numerically evolving the electroweak equations of motion on a
lattice. Debris from sphaleron decay was analysed for magnetic helicity
and the connection between baryon number violation and generation of
magnetic helicity was confirmed. The aim of the present paper is to
{\it analytically} study the sphaleron decay process and to derive the
generation of magnetic helicity in a transparent manner.

Although the full dynamics of sphaleron decay involves the complete
electroweak equations of motion, we are able to extract the dynamics
relevant to the electromagnetic sector following a scheme similar
to Ref.~\cite{Klinkhamer:1984di}.
Our strategy is to first consider the ``$SU(2)$ sphaleron''
which is an explicitly known saddle-point solution in the electroweak
model with the hypercharge coupling, $g'$, set to zero \cite{Manton:1983nd}.
Not only is the $SU(2)$ sphaleron known explicitly but a path in field
space that connects the sphaleron solution to the vacuum solution has
also been constructed \cite{Manton:1983nd}. Let $\mu$ be the parameter
along this path. We will consider the decay of the sphaleron along this
particular path in field space.
Hence the time dependence of the fields is given by the (unspecified)
function $\mu(t)$. With this model of the decaying sphaleron, we
now turn on the hypercharge coupling, $g'$, assuming it to be
small. Then the decaying sphaleron produces electromagnetic
currents, which produce electromagnetic fields. We calculate
the magnetic helicity in these fields and show analytically
that it is conserved at late times and asymptotes to some
non-zero value which depends on $\mu(t)$. We can also show
that the sign of the magnetic helicity depends on which side
of the saddle the sphaleron rolls down. Thus the sign of the
magnetic helicity is directly related to whether baryons
or antibaryons are produced during the decay of the sphaleron.
Since we know excess baryons are produced in the universe,
cosmic magnetic fields carry left-handed magnetic helicity.
Furthermore, the magnitude of the helicity is directly proportional
to the cosmic baryon number density as discussed in
\cite{Cornwall:1997ms,Vachaspati:2001nb,Copi:2008he}.

\section{$SU(2)$ Sphaleron}
\label{sec:su2sphaleron}

The $SU(2)$ sphaleron is a saddle-point solution to the static
electroweak equations with the hypercharge coupling, $g'$, set
to zero. For our purpose, it is necessary to consider a path
of field configurations that starts with the sphaleron and
ends at the vacuum. Such a path was constructed by Manton
\cite{Manton:1983nd} in order to implement a Morse theory
argument indicating the existence of a sphaleron solution. The
path in field space is now written in spherical coordinates,
$(r,\theta,\phi)$, and $\mu$ is a parameter along the path
\begin{eqnarray}
\Phi (\mu,r,\theta,\phi) &=& \frac{v}{\sqrt{2}} \biggl [
                 (1-h(r)) \begin{pmatrix}
                        0\\ e^{-i\mu}\cos\mu
                      \end{pmatrix}
     + h(r) \Phi^\infty (\mu,\theta,\phi) \biggr ]
                     \nonumber \\
W_i (\mu,r,\theta,\phi) &=& i \frac{2}{g} f(r)
                             W^\infty_i (\mu,\theta,\phi)
\label{PhiW}
\end{eqnarray}
where $g$ is the $SU(2)$ coupling, $v/\sqrt{2}$ is the vacuum expectation
value of the Higgs field, $h(r)$ is the Higgs profile function
and $f(r)$ the gauge profile function.

The asymptotic fields are given by
\begin{equation}
\Phi^\infty (\mu,\theta,\phi) = \begin{pmatrix}
            \sin\mu ~ \sin\theta ~ e^{i\phi}\\
   e^{-i\mu} (\cos\mu + i\sin\mu ~ \cos\theta )
                                 \end{pmatrix}
\end{equation}
\begin{equation}
W^\infty_i = - \partial_i U^\infty ~ (U^\infty)^{-1}
\label{Winftyi}
\end{equation}
where
\begin{equation}
U^\infty = \left ( \begin{array}{cc}
         \Phi_2^{\infty *} & \Phi_1^\infty \\
        -\Phi_1^{\infty *} & \Phi_2^\infty
            \end{array} \right ) \ .
\end{equation}

The sphaleron solution is obtained if we set $\mu =\pi/2$ and the vacuum
is obtained for $\mu=0$ or $\mu=\pi$. The path from $\mu=\pi/2$ to
$\mu=0$ describes the sphaleron rolling down one side of the saddle,
and the path from $\mu=\pi/2$ to $\mu=\pi$ describes roll down on the
other side of the saddle. The former path describes the creation of
antibaryons while the latter describes the creation of baryons.

The dimensionless sphaleron profile functions are found by numerically
solving the static electroweak equations of motion. However, the essential
features of the profiles are captured by the simple ansatz called
``Ansatz {\it a}'' in \cite{Klinkhamer:1984di},
\begin{eqnarray}
f^a (r) = \begin{cases}
          ( r/r_f )^2, & \mbox{if }
                                              \mbox{ $r < r_f$}\\
                  1, & \mbox{if } \mbox{ $r \ge r_f$}
              \end{cases}
\label{fansatz}
\end{eqnarray}
\begin{eqnarray}
h^a (r ) = \begin{cases}
                {r}/{r_h} ,& \mbox{if } \mbox{ $r < r_h$} \\
                1, & \mbox{if } \mbox{ $r \ge r_h$}
              \end{cases}
\label{hansatz}\ .
\end{eqnarray}
The length scales $r_f$ and $r_h$ are related to
the vector and scalar masses in the model. For simplicity
we will take $r_f=r_h$,  when we numerically evalute
certain integrals in Sec.~\ref{sec:timedependence}.

\section{Electromagnetic currents}
\label{sec:currents}

To model the decay of the sphaleron, we take the parameter
$\mu$ to be a function of time, $\mu(t)$, with $\mu(0)=\pi/2$
and $\mu(\infty)=\pi$. The Higgs and $SU(2)$ gauge fields are
still given by Eqs.~(\ref{PhiW}).
As discussed in \cite{Klinkhamer:1984di}, we can now consider
a small hypercharge coupling constant, $g'$, and evaluate the
electromagnetic field perturbatively in $g'$. Then the decaying
sphaleron will be a source for an electromagnetic field. If
we denote the electromagnetic gauge field by $A_\mu$, Maxwell
equations in the Lorenz gauge give
\begin{equation}
(\partial_t^2 - {\bf \nabla}^2) A^\mu = J^\mu \ ,
\end{equation}
where $J^\mu$ is the electromagnetic current produced by the
decaying sphaleron. To leading order in the hypercharge coupling
constant, $g'$, the electromagnetic current is equal to the
hypercharge current given by
\begin{equation}
J^\mu = - i \frac{g'}{2} \left [ \Phi^\dag D^\mu \Phi -
                                (D^\mu \Phi )^\dag \Phi \right ]\ .
\label{Jmu}
\end{equation}

To calculate the magnetic helicity produced by sphaleron decay
we only need to find the spatial components of $A_\mu$. Hence
we only need the spatial components of the current $J^\mu$.
Eq.~(\ref{Jmu}) gives the contravariant current vector components;
we prefer to work with the ``ordinary'' components (see Sec.~4.8 of
\cite{Weinberggrav}) that are obtained by dividing by the square
root of the metric factors, namely, $r$ and $r\sin\theta$ in the
$\theta$ and $\phi$ components. The ordinary components of the
current are
\begin{eqnarray}
J_r &=&  \frac{g'v^2}{4} \cos\theta ~ h'(r) \sin(2\mu)\nonumber \\
    &\equiv&  \frac{g'v^2}{4} \cos\theta ~ {\cal J}_r (t,r)
\label{Jr}\\
J_\theta &=&  \frac{g'v^2}{4}  \sin\theta ~ \frac{\sin (2\mu)}{r} \left [
     (1-h)f \cos^2\mu - h(1-f)+hf(1-h)~\sin^2\mu \right ]
\nonumber \\
         &\equiv& \frac{g'v^2}{4} \sin\theta ~ {\cal J}_\theta (t,r)
\label{Jtheta}\\
J_\phi &=&  \frac{g'v^2}{4} \sin\theta ~ \frac{2\sin^2\mu}{r} \left [
      f(1-h)^2\cos^2\mu + h^2 (1-f) \right ] \nonumber \\
         &\equiv&  \frac{g'v^2}{4} \sin\theta ~ {\cal J}_\phi (t,r)\ ,
\label{Jphi}
\end{eqnarray}
where the prime on $h$ denotes differentiation with respect to $r$.

The spherical components of the current, $J_i$, as given in
Eq.~(\ref{Jmu}) only involve spatial derivatives of $\Phi$ and the
spacelike components of the gauge fields. Hence the three-current does
not involve time derivatives of $\mu(t)$ or the time-component of the
gauge fields. Also, the zeroth (time) component of the current is
unspecified since we will not need it in our calculation. We assume
it is such as to ensure current conservation.

The angular form of the currents is of interest since
there is an azimuthal component, $J_\phi$, and the
$J_r,~J_\theta$ components circulate in the constant $\phi$ plane.
Hence the $J_\phi$ current resembles circular
current-carrying wires in planes parallel to the $xy-$plane
and the $J_r, ~J_\theta$ components resemble current-carrying
wires wrapped on a toroidal solenoid. The two components of the
current are linked and hence the current is helical as measured by
${\bf J}\cdot {\bf \nabla}\times {\bf J}$.

Already we can see the relation of the sign of the helicity and
the decay path taken by the sphaleron. The path from $\mu=\pi/2$
to $\mu=\pi$ is related to the path from $\mu=\pi/2$ to $\mu=0$
by the transformation $\mu \to \pi-\mu$. Under this transformation
$J_\phi$ is unchanged but $J_r$ and $J_\theta$ change sign.
Thus the linking number of the current components is reversed.
This implies that the magnetic helicity produced by the currents
will also be reversed.

We will eventually need to Fourier transform the current. For this we
find it most convenient to work with Cartesian components
\begin{eqnarray}
J_x (t,{\bf x}) &=& J_r \sin\theta ~\cos\phi +
                  J_\theta \cos\theta ~\cos\phi - J_\phi \sin\phi
\label{Jx} \\
J_y (t,{\bf x}) &=& J_r \sin\theta ~\sin\phi +
                  J_\theta \cos\theta ~\sin\phi + J_\phi \cos\phi
\label{Jy} \\
J_z (t,{\bf x}) &=& J_r \cos\theta - J_\theta \sin\theta
\label{Jz}\ .
\end{eqnarray}
The Fourier transform is implemented by using the plane wave
expansion (see Sec.~11.3 of \cite{MorFes53})
\begin{equation}
e^{i{\bf k}\cdot {\bf x}} = \sum_{n=0}^\infty (2n+1)i^n
     \sum_{m=0}^n \epsilon_m \frac{(n-m)!}{(n+m)!}
     \cos(m(\phi-v)) P_n^m (\cos u) P_n^m (\cos\theta) j_n(kr)\ ,
\label{planewave}
\end{equation}
where
\begin{eqnarray}
{\bf x}&=& r(\sin \theta \cos \phi , \sin \theta \sin \phi, \cos \theta)
\nonumber \\
{\bf k}&=& k(\sin u \cos v , \sin u \sin v, \cos u) \ ,
\end{eqnarray}
and $j_n(kr)$ is the spherical Bessel function of order $n$,
$P_n^m(\cdot )$ is the associated Legendre function,
and $\epsilon_0=1$, $\epsilon_m =2$ for $m \ge 1$.

Now, for example, the Fourier transform of the $x-$component of
the current is
\begin{equation}
{\tilde J}_x (t,{\bf k}) =
     \int d\Omega \int_0^\infty dr ~r^2  ~
         e^{i{\bf k}\cdot {\bf x}} J_x (t,{\bf x})\ .
\end{equation}
In this we insert $e^{i{\bf k}\cdot {\bf x}}$ from Eq.~(\ref{planewave})
and the current from (\ref{Jx}), using (\ref{Jr})-(\ref{Jphi}).
After quite a bit of algebra, we find
\begin{eqnarray}
{\tilde J}_x (t,{\bf k}) &=& -\frac{g'v^2}{4} 4\pi \sin u ~
                       [ \cos u \cos v ~ K_2 (t,k)+i\sin v ~ K_1(t,k)]
      \label{Jkx} \\
{\tilde J}_y (t,{\bf k}) &=& -\frac{g'v^2}{4} 4\pi \sin u ~
                       [ \cos u \sin v ~ K_2 (t,k)-i\cos v ~ K_1(t,k)]
      \label{Jky} \\
{\tilde J}_z (t,{\bf k}) &=&  \frac{g'v^2}{4} \frac{4\pi}{3}
                       [~ K_0(t,k) + (1-3\cos^2u) ~ K_2 (t,k)]\ ,
       \label{Jkz}
\end{eqnarray}
where
\begin{eqnarray}
K_0 &=& \int_0^\infty dr r^2 j_0 (kr)
          [{\cal J}_{r}(t,r)-2{\cal J}_{\theta}(t,r)]
    \nonumber \\
K_1 &=& \int_0^\infty dr r^2 j_1 (kr) {\cal J}_\phi(t,r)
    \nonumber \\
K_2 &=& \int_0^\infty dr r^2 j_2 (kr) {\cal J}_{r+\theta}(t,r)\ ,
\end{eqnarray}
and
\begin{equation}
{\cal J}_{r+\theta}(t,r) \equiv {\cal J}_r(t,r) + {\cal J}_\theta (t,r) \ .
\end{equation}

\section{Magnetic helicity}
\label{sec:helicity}

The magnetic helicity is defined by
\begin{equation}
{\cal H}(t) = \int d^3 x ~ {\bf A} \cdot {\bf B}\ .
\label{Ht}
\end{equation}

The procedure to evaluate magnetic helicity is now straightforward
though cumbersome. We find ${\bf A}$ from Maxwell's equation, then
the magnetic field, ${\bf B} = {\bf \nabla}\times {\bf A}$, and finally
the resulting magnetic helicity.
In terms of the fields in Fourier space (denoted by over tilde's)
\begin{equation}
{\tilde {\bf A}} (k) = - \frac{{\tilde {\bf J}} (k)}{\omega^2 -{\bf k}^2}\ ,
\label{Ak}
\end{equation}
where ${\tilde {\bf J}}(k)$ is given in Eqs.~(\ref{Jkx})-(\ref{Jkz})\ ,
and
\begin{equation}
{\tilde {\bf B}} (k) = -i{\bf k} \times {\bf A} (k)\ .
\label{Bk}
\end{equation}
For us it is more useful to work with the three dimensional (spatial)
Fourier transform
of the fields while not transforming their time dependence. This is
because the time dependence occurs via the unknown function $\mu (t)$.
To go from the frequency dependent function to the time dependent
function, we use
\begin{equation}
\int_{-\infty}^{+\infty} d\omega
   \frac{e^{i\omega (t-\tau)}}{\omega^2-{\bf k}^2}
 = \frac{2\pi \sin(k(t-\tau))}{k} \Theta (t-\tau)
\label{freqint}\ ,
\end{equation}
where we have chosen a contour prescription so as to get retarded
solutions.

We use Eqs.~(\ref{Ak}) and (\ref{Bk}) in the helicity integral
(\ref{Ht}) and then integrate over the frequencies using
(\ref{freqint}). The result is
\begin{eqnarray}
{\cal H}(t) = i \int \frac{d^3k}{(2\pi)^3}
\int_{-\infty}^t d\tau \int_{-\infty}^t d\tau'
\frac{\sin(k(t-\tau ))}{k} \frac{\sin(k(t-\tau' ))}{k}
{\bf k}\cdot {\tilde {\bf J}}(\tau, -{\bf k}) \times
             {\tilde {\bf J}}(\tau ', {\bf k})\ .
\end{eqnarray}
The triple product can be evaluated using the explicit expressions
for the current components in Eqs.~(\ref{Jkx})-(\ref{Jkz}).

Further simplification involves considerable algebra and integration
over trigonometric functions. We also use the spherical Bessel function
relation
\begin{equation}
j_2 (x) = \frac{3}{x}j_1 (x) - j_0(x)\ ,
\end{equation}
and perform the integrations over the angular coordinates $u,~v$ in
$k-$space to eventually get
\begin{equation}
{\cal H} (t) = \frac{16}{3} \left ( \frac{g'v^2}{4} \right )^2
               \left [ {\cal H}_0 (t) + {\cal H}_1(t) \right ],
\end{equation}
where
\begin{eqnarray}
{\cal H}_0 (t) &=&- \int_{-\infty}^t d\tau \int_{-\infty}^t d\tau'
\int_0^\infty dr ~r^2 \int_0^\infty dr' ~r'^2
{\cal J}_\phi (\tau,r) {\cal J}_{\theta}(\tau',r')
\nonumber \\
&& \times \int_0^\infty dk~k \left [ \cos (k(\tau -\tau')) -
        \cos (k (2t-\tau-\tau')) \right ] j_1(kr)j_0(kr')
\end{eqnarray}
\begin{eqnarray}
{\cal H}_1 (t) &=& \int_{-\infty}^t d\tau \int_{-\infty}^t d\tau'
\int_0^\infty dr ~r^2 \int_0^\infty dr' ~r'^2
{\cal J}_\phi (\tau,r) {\cal J}_{r+\theta}(\tau',r')
\nonumber \\
&& \times \frac{1}{r'} \int_0^\infty dk \left [ \cos (k(\tau -\tau')) -
      \cos (k (2t-\tau-\tau')) \right ] j_1(kr) j_1(kr') \ .
\end{eqnarray}

The integrations over $k$ are performed using the explicit
expressions for the spherical Bessel functions in terms of
trigonometric functions
\begin{equation}
j_0(x) = \frac{\sin(x)}{x}
\end{equation}
\begin{equation}
j_1(x) = \frac{\sin(x)}{x^2} - \frac{\cos(x)}{x}\ .
\end{equation}
We find
\begin{equation}
\int_0^\infty dk ~k ~\cos(k\delta) j_1(kr) j_0(kr') =
\frac{\pi}{8r^2 {r'}^2} P_0(\delta ,r,r')
\end{equation}
\begin{equation}
P_0(\delta,r,r')= r' [ S_0(\delta,r,r'
)-S_0(\delta,-r,r')-S_0(\delta,r,-r')+S_0(\delta,-r,-r')]\ ,
\label{P0}
\end{equation}
with
\begin{equation}
S_0(\delta,r,r')= (\delta+r')~ {\rm sign}(\delta+r+r')
\label{S0}
\end{equation}
and
\begin{equation}
\int_0^\infty dk ~\cos(k\delta) j_1(kr) j_1(kr') =
\frac{-\pi}{48 r^2 {r'}^2} P_1(\delta,r,r')\ ,
\end{equation}
where
\begin{equation}
P_1(\delta,r,r') = S_1(\delta,r,r')-S_1(\delta,-r,r')-S_1(\delta,r,-r')+S_1(\delta,-r,-r')\ ,
\label{P1}
\end{equation}
with
\begin{equation}
S_1(\delta,r,r')= ( \delta^3 - 3(r^2+r'^2)\delta - 2(r^3+r'^3) ) ~ {\rm sign}(\delta+r+r')\ .
\label{S1}
\end{equation}

With these integrations, the final expressions for ${\cal H}_0$
and ${\cal H}_1$ are
\begin{equation}
{\cal H}_0 (t) = -\frac{\pi}{8} \int_{-\infty}^t d\tau
\int_{-\infty}^t d\tau' \int_0^\infty dr \int_0^\infty dr'
{\cal J}_\phi(\tau,r){\cal J}_{\theta}(\tau',r')
\left [ P_0(\tau-\tau',r,r') - P_0(2t-\tau-\tau',r,r') \right ]
\label{calH0final}
\end{equation}
\begin{equation}
{\cal H}_1 (t) = -\frac{\pi}{48} \int_{-\infty}^t d\tau
\int_{-\infty}^t d\tau' \int_0^\infty dr \int_0^\infty \frac{dr'}{r'}
{\cal J}_\phi(\tau,r) {\cal J}_{r+\theta}(\tau',r')
\left [
P_1(\tau-\tau',r,r') - P_1(2t-\tau-\tau',r,r')
\right ]\ .
\label{calH1final}
\end{equation}

\section{Asymptotic magnetic helicity}
\label{sec:asymptotic}

In the integrations for ${\cal H}_0(t)$ and ${\cal H}_1(t)$
in Eqs.~(\ref{calH0final}) and (\ref{calH1final}), the factors
${\cal J}_r$, ${\cal J}_\theta$ and ${\cal J}_\phi$ have
compact support. So the integrand is non-vanishing for some
finite domain of $\tau$, $\tau'$, $r$ and $r'$. If we now
take $t$ to be large, while keeping $\tau$, $\tau'$, $r$
and $r'$ finite, then $2t-\tau-\tau'$ is large and also
large compared to $r$ and $r'$. Then the ``sign'' factors
in Eqs.~(\ref{S0}) and (\ref{S1}) can be replaced by $+1$
and it can be checked that
$P_0(2t-\tau-\tau',r,r')=0$ and $P_1(2t-\tau-\tau',r,r')=0$.
Hence at late times, only the time independent pieces in
${\cal H}_0$ and ${\cal H}_1$ can survive, and we get
\begin{equation}
{\cal H}_0 (\infty ) = - \frac{\pi}{8} \int_{-\infty}^\infty
d\tau \int_{-\infty}^\infty d\tau' \int_0^\infty dr \int_0^\infty dr'
{\cal J}_\phi(\tau,r) {\cal J}_{\theta}(\tau',r')
P_0(\tau-\tau',r,r')
\end{equation}
\begin{equation}
{\cal H}_1 (\infty) = -\frac{\pi}{48} \int_{-\infty}^\infty
d\tau \int_{-\infty}^\infty d\tau'
\int_0^\infty dr \int_0^\infty \frac{dr'}{r'}
{\cal J}_\phi(\tau,r) {\cal J}_{r+\theta}(\tau',r')
P_1(\tau-\tau',r,r')
\end{equation}
and
\begin{equation}
{\cal H}(\infty)= \frac{16}{3} \left ( \frac{g'v^2}{4} \right )^2
           \left [ {\cal H}_0(\infty) + {\cal H}_1(\infty) \right ]\ .
\end{equation}

In terms of the mass of the $W$-boson, $m_W = gv/2$, the helicity
can be written as
\begin{equation}
{\cal H}(\infty)= \frac{16}{3} \frac{\sin^2\theta_w}{g^2} m_W^4
           \left [ {\cal H}_0(\infty) + {\cal H}_1(\infty) \right ]\ ,
\end{equation}
where $\theta_w$ is the weak mixing angle and for $g' \ll g$,
$\sin\theta_w \sim \theta_w \sim g'/g$.
The quantities ${\cal H}_0$ and ${\cal H}_1$ have dimensions of
$L^4$ where $L$ is a length scale that enters the profile functions.
Assuming $r_f \sim r_h$ in the profile functions and left-handed
helicity (see sections below),
the parametric dependence of the helicity is
\begin{equation}
{\cal H}(\infty)\sim - \frac{\sin^2\theta_w}{g^2} (m_W r_f )^4\ .
\end{equation}
The result can also be expressed in terms of the fine structure
constant, $\alpha = e^2/4\pi \approx 1/137$, using
$e=\sqrt{g^2+g'^2}\sin\theta_w\cos\theta_w$.

The asymptotic helicity depends on the time evolution chosen
in $\mu(t)$ and also on the profile functions. There is no
apparent symmetry reason for the integrals to vanish, and hence
the decay of the sphaleron will lead to a magnetic field with
non-vanishing magnetic helicity. Below we will numerically
evaluate the time-dependent magnetic helicity for a choice of
$\mu (t)$ and the ansatz for the sphaleron profile functions in
Eqs.~(\ref{fansatz}) and (\ref{hansatz}).
This evaluation also gives the asymptotic numerical value of
the magnetic helicity.

\section{Time dependent magnetic helicity}
\label{sec:timedependence}

Eqs.~(\ref{calH0final}) and (\ref{calH1final}) give the time
dependent magnetic helicity but the integrations are difficult
to do analytically, especially because the true profile functions
are not known in closed form, and the time evolution function,
$\mu(t)$, needs to be chosen. Even with the very simple profile
Ansatz of Eqs.~(\ref{fansatz}) and (\ref{hansatz}), the integrations
lead to a lot of different terms.
So we have evaluated the integrals numerically but this
also requires a choice of the time dependent function, $\mu(t)$.

One possible scheme to find a suitable $\mu(t)$ is to derive an effective
equation of motion by substituting the field ansatz for the $SU(2)$
sphaleron in Sec.~\ref{sec:su2sphaleron} in the energy functional.
This evaluation was done in Ref.~\cite{Manton:1983nd} by taking
$\mu$ to be a parameter and not a dynamical variable. The result
was a ``potential energy'' function
\begin{equation}
V(\mu) = m_W (C_1 \sin^2\mu + C_2 \sin^4\mu)\ ,
\end{equation}
where $C_1$ and $C_2$ are numerical constants. If we promote $\mu$
to a dynamical variable, the energy will also depend on ${\dot \mu}$
and we can obtain an equation of motion for $\mu$. However, this exercise
is moot because it leads to a conservative equation of motion for
$\mu$, while the true evolution is dissipative, and once $\mu$
gets to the true vacuum value ($0$ or $\pi$) it should stay there.
One way to introduce dissipation is to replace second time derivatives
by first time derivatives in the equation of motion. This suggests
an equation of motion
\begin{equation}
{\dot \mu} = A_1 \sin\mu + A_2 \sin^3\mu \ ,
\end{equation}
where $A_1$ and $A_2$ are some coefficients. The equation can be
integrated and its messy solution is very similar to the much
simpler
\begin{equation}
\mu(t) = \frac{\pi}{2}\left [ 1 + \tanh \left (\frac{t}{r_t}
                         \right ) \right ] \ .
\end{equation}
Hence we adopt this functional form for $\mu(t)$ and proceed with
our numerical evaluation of the magnetic helicity. Note that
$\mu (0)=\pi /2$ and $\mu (\infty) = \pi$, and hence this choice
of $\mu (t)$ describes creation of baryons.

\begin{figure}
  \includegraphics[height=0.45\textwidth,angle=0]{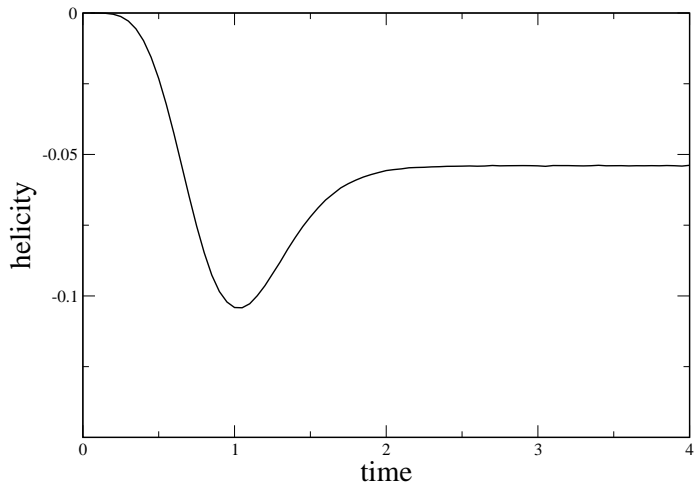}
  \caption{${\cal F}(t)$ versus $t/r_t$ for $r_t=r_f=r_h$.
}
\label{helicityevolution}
\end{figure}

Next we write
\begin{equation}
{\cal H} (t) = \frac{\sin^2\theta_w}{g^2} (m_W r_f )^4 {\cal F}(t) \ ,
\end{equation}
with the dimensionless function ${\cal F}(t)$ defined as
\begin{equation}
{\cal F}(t) = \frac{16}{3 r_f^4}
               \left [ {\cal H}_0 (t) + {\cal H}_1(t) \right ] \ .
\end{equation}
The numerically evaluated ${\cal F}(t)$ vs. $t$ is shown in
Fig.~\ref{helicityevolution} for $r_h=r_f=r_t$. An interesting
feature of the plot is that the magnetic helicity is negative
and conserved at late times.

The plot in Fig.~\ref{helicityevolution} corresponds to Fig.~1 of
Ref.~\cite{Copi:2008he} where it was obtained by solving the
electroweak field equations on a spatial lattice starting with
a perturbed sphaleron.

\section{Sign of helicity}
\label{sign}

An important virtue of the present calculation is that it can relate the
sign of the asymptotic magnetic helicity to the direction in which the
sphaleron decays, which in turn is also related to whether baryons or
antibaryons are produced. So far we have only seen this connection by
numerically evaluating the helicity integrals in the limit
$t \to \infty$. Here we will analytically establish the connection for
small times, just as the sphaleron starts to decay.

The sphaleron is located at $\mu=\pi/2$. Therefore we consider
\begin{equation}
\mu = \frac{\pi}{2} + \epsilon (t) = \frac{\pi}{2}
             + {\dot \epsilon}_0 \delta t + O((\delta t)^2) \ ,
\end{equation}
where $\delta t$ is a small time interval and ${\dot \epsilon}_0$
denotes the velocity with which the sphaleron starts to roll.
Now we can expand ${\cal J}_r$, ${\cal J}_\theta$ and ${\cal J}_\phi$
to lowest order in $\delta t$. This gives
\begin{equation}
{\cal J}_{r} \sim  F_r(r) {\dot \epsilon}_0 \delta t
                            + O((\delta t)^2)
\label{deltaJr}
\end{equation}
\begin{equation}
{\cal J}_{\theta} \sim  F_\theta (r) {\dot \epsilon}_0 \delta t
                            + O((\delta t)^2)
\label{deltaJt}
\end{equation}
\begin{equation}
{\cal J}_\phi = F_\phi (r) + O((\delta t)^2) \ ,
\end{equation}
where $F_r$, $F_\theta$ and $F_\phi$ are some time-independent functions.
Next we count powers of $\delta t$ in the helicity integrals. We
get two powers from the two time integrals, one from the factors of
the currents as given above, and one power from the
terms in square brackets in Eqs.~(\ref{calH0final}) and
(\ref{calH1final}). Therefore
\begin{equation}
{\cal H}(\delta t) \sim - {\dot \epsilon}_0 (\delta t)^4 \times
                        {\cal I} \ ,
\end{equation}
where ${\cal I}$ is an integral that does not depend on $\mu (t)$.
Since ${\cal I}$ still depends on the profile functions, we can
only evaluate it numerically and, as is clear from
Fig.~\ref{helicityevolution}, ${\cal I} > 0$.
This tells us that an increase of $\mu$ (${\dot \epsilon}_0 >0$),
corresponding to baryon production,
yields left-handed magnetic helicity, and a decrease of $\mu$
(${\dot \epsilon}_0 < 0$)
leads to right-handed magnetic helicity. In the cosmological
context, since we know that baryons prevail, this predicts that
left-handed magnetic helicity is prevalent in the universe
\cite{Vachaspati:2001nb}.

\section{Conclusions}
\label{sec:conclusions}

We have developed an analytical technique to study the magnetic
debris produced during baryon number violating processes that
occur via sphalerons. We have calculated the time-dependent helicity
of the magnetic fields that are produced and found the asymptotic
helicity in terms of the decay path of the sphaleron up to
quadrature. The sign of
the helicity depends on whether baryons or antibaryons are produced
and baryon production leads to left-handed magnetic helicity.
Our final result for the time evolution of magnetic helicity is shown
in Fig.~\ref{helicityevolution} and compares well with the numerical
computations of Ref.~\cite{Copi:2008he}. These figures
demonstrate that magnetic helicity is conserved at late times.
This is novel because the conservation of magnetic helicity
is usually discussed in the magneto-hydrodynamic context where
a highly conducting plasma is present. In our case there is
no external plasma. The (approximate) conservation of helicity
has also been noted in solutions to the {\it vacuum} Maxwell
equations in Ref.~\cite{Jackiw:1999bd}.

The wider implication of our results is that there is likely
to be a unified origin of cosmic matter and cosmic magnetic
fields. This possibility has been discussed before
\cite{Vachaspati:1991nm,Vachaspati:1994xc,Vachaspati:1994ng,
Cornwall:1997ms,Vachaspati:2001nb, Copi:2008he}. The analytical
technique we have developed makes
the connection between magnetic helicity and baryon number more
transparent and has the potential for further development.
For example, one could consider an ensemble of possible
decay paths described by the function $\mu(t)$, and the mean
magnetic helicity could then be evaluated as an average over
this ensemble. Further one could incorporate our results in a
realistic baryogenesis scenario and derive spectral properties of
the magnetic field. A simple analysis of this kind was done in
Ref.~\cite{Ng:2010mt}, where a model for the magnetic fields produced
during sphaleron decay was assumed. Based on our present results,
it appears that the model captures most of the features of the
magnetic fields produced in sphaleron debris, and so it is worth
re-displaying the magnetic field expressions here. In spherical
coordinates with origin at the location of the sphaleron, the model
for the magnetic field is
\begin{eqnarray}
B_r &=&  - \frac{a \cos\theta}{(a^2+r^2)^{3/2}} \nonumber \\
B_\theta &=& + \frac{a \sin\theta}{2}
            \frac{2a^2-r^2}{(a^2+r^2)^{5/2}}
\label{Bchoice}                                         \\
B_\phi &=& \frac{r}{a^3} e^{-r/a} \sin\theta \nonumber \ .
\end{eqnarray}
The extent of the magnetic field is determined by the length
scale $a$, which is time-dependent. The simplest choice is
\begin{equation}
a(t) = t-t_0 \ ,
\end{equation}
where $t_0$ is the time at which the sphaleron started rolling
down. As the magnetic field expands, just as in the case of the
sphaleron, the helicity stays constant at
\begin{equation}
{\cal H} = - \frac{8\pi}{3} 0.57 = - 4.8 \ .
\end{equation}
The numerical value of the asymptotic helicity can be adjusted by
suitably rescaling the magnetic field.

Once a space-filling helical magnetic field has been produced at
the baryogenesis epoch, further evolution will be subject to
magneto-hydrodynamical evolution in a cosmological setting
e.g. \cite{Vachaspati:2001nb,Jedamzik:2010cy}. We expect the presence
of magnetic helicity to have significant impact on the evolution.
This topic, as well as the possibility of observation of early
magnetic fields, deserves further investigation.

\begin{acknowledgments}
This work was supported by the DOE at Arizona State University.
\end{acknowledgments}



\end{document}